\documentclass[aps,preprintnumbers,secnumarabic,amsmath,amssymb,usenatbib,nofootinbib,superscriptaddress,showkeys,showpacs,11pt]{revtex4-2}
\usepackage{amsmath}
\usepackage{amsfonts,color}
\usepackage{amssymb,float}
\usepackage{graphicx}
\usepackage{appendix}
\usepackage[colorlinks]{hyperref}
\usepackage{xcolor}
\usepackage{orcidlink}
\usepackage{epsf}
\usepackage{bm}
\usepackage{epstopdf}
\usepackage{natbib}
\usepackage{lipsum}

\begin{document}

\title{The interaction of extended Bose-Einstein condensate dark matter with viscous $f(T, B)$  gravity}

\author{E. Mahichi}
\email{elhammahichi@gmail.com}
\affiliation{Department of Physics, Ayatollah Amoli Branch, Islamic Azad University, Amol, Iran}

\author{Alireza Amani\orcidlink{0000-0002-1296-614X}}
\email{ Corresponding Author: al.amani@iau.ac.ir}
\affiliation{Department of Physics, Ayatollah Amoli Branch, Islamic Azad University, Amol, Iran}

\date{\today}


\begin{abstract}

In this paper, we study the viscous $f(T, B)$ gravity model as a source of dark energy, and the Extended Bose-Einstein Condensate (EBEC) as a source of dark matter, in a flat-FRW metric. In the presence of bulk viscosity, we obtain Friedmann equations and write two continuity equations of dark energy and dark matter by interacting them. Using the generalized Gross-Pitaeveskii equation, we earn Equation of State (EoS) of dark matter by EBEC regime as $p_m = \alpha \rho_{m} + \beta \rho_{m}^2$ in which the both of terms are respectively introduced as normal dark matter and dark matter halo. The innovation of the work is that we can simultaneously describe the nature of the dark parts of the universe with the viscous $f(T, B)$ gravity and the EBEC regime, which leads to a deep understanding of the different epochs of the universe from early to late times. In what follows, the energy density and the pressure of dark energy are reconstructed in terms of the redshift parameter, and then we fit the obtained results with 53 supernova data from the Hubble data constraints. Next, we plot the cosmological parameters in terms of the redshift parameter and conclude that the current universe is in an accelerated phase. Finally, we analyze the stability and instability of the current model with the sound speed parameter as well as we draw the density parameter values for dark energy in terms of the redshift parameter.

\end{abstract}
\maketitle
\newpage


\section{Introduction}\label{I}

One of the greatest concerns of human beings is to know about the universe, which requires knowing the contents of the universe, such as dark energy, dark matter, and so on. From the very beginning, many efforts were made to this end, including Friedmann and his famous equations, although Hubble's contribution to this issue is also important because Hubble-Lumiter's law confirmed that the universe is expanding at a constant speed. This velocity is proportional to the distance of the galaxies from the Earth. Because of the importance of knowing the universe, in recent decades, by measuring cosmic distances and observing cosmic objects beyond the Milky Way, they have found that the universe is expanding rapidly. Therefore, the expansion of the universe at a non-constant speed was discovered and confirmed by Ia-type supernovae, cosmic microwave background radiation, and large-scale structures in the universe  \cite{Riess_1998, Perlmutter_1999, Bennett_2003, Tegmark_2004}. The cause of this event is a mysterious and unknown force that is greater than the force of gravity but in the opposite direction, which is called dark energy. Since the universe is considered a thermodynamic system, and its evolution depends only on the internal contents of the universe, therefore, due to the lack of matter-energy transfer from outside to inside, must be attributed to a strong negative pressure that acts as a repulsion that leads to the expansion of the universe. With these explanations, the problem of dark energy can be interpreted in the fundamental theoretical framework of string theory and quantum gravity. Extensive studies on dark energy with models such as the cosmological constant \cite{Weinberg-1989, Ng-1992}, scalar fields \cite{Chiba-2000, Kamenshchik-2001, Caldwell-2002, Singh-2003,Guo-2005, Wei-2005, Setare1-2009, Amani-2011, Sen-2002, Bagla-2003, Sadeghi1-2009, Setare-2009, Amani-2013, Amani-2014, Battye-2016, Li-2012}, modified gravity \cite{Capozziello-2002, Dolgov-2003, Copeland-2006, Faraoni-2006, Nojiri-2007, Felice-2010, CapozzielloS-2011, Iorio-2016, Nojiri-2017}, holography and agegraphic \cite{Li-2004, Wei-2009, Amani1-2011, Campo-2011, Hu-2015, Wei-2007, Jamil-2010, Jawad-2013}, bouncing theory \cite{Shtanov-2003, Sadeghi-2009, Sadeghi-2010, Amani-2016, Singh-2016}, teleparallel gravity \cite{Capozziello-2011, Myrzakulov-2011, Pourbagher_2019, Rezaei-2017, pourbagher1-2020} and braneworld models \cite{Sahni-2003, Setare-2008, Brito-2015}.

Among these mentioned models, teleparallel gravity is considered to be one of the suitable candidates for dark energy, which was first introduced by Einstein for the union between electromagnetism and gravity. This theory states that space-time is associated with a curvature-free linear relationship by a metric tensor field called tetrad-vector field dynamics. In that case, a tetrad field is naturally used to define a linear Weitzenböck connection, which represents a torsion connection without curvature. But Levi-Civita connection is used in the geometry of general relativity to show the curvature connection without torsion. Therefore, by converting the tetrad components and metric components to each other, we can convert teleparallel gravity model to the general relativity model and vice versa. This means, the curvature term in general relativity changes to a torsion term in the teleparallel model and vice versa, which confirms the claim that teleparallel gravity and general relativity are equivalent. This equivalent is caused the relation $R = -T + 2 \nabla_\mu T^\mu$ in which $R$, $T$, and $T^\mu$ are curvature scalar, torsion scalar, and torsion vector, respectively. Nevertheless, the corresponding field equations are clearly covariant and then the theory has a local Lorentz invariant \cite{Li-2011}. By converting $T$ to $f(T)$, we expand teleparallel gravity to modified teleparallel gravity, where $f(T)$ is the function of the torsion scalar. Also, by introducing $B = 2 \nabla_\mu T^\mu$, we can redevelop the modified teleparallel gravity in the form $f(T, B)$ gravity in which $B$ is the boundary term. Thus, model $f(T, B)$ gravity can simultaneously covers both the models of $f(T)$ gravity and $f(R)$ gravity with help of relation $R = -T + B$, i.e., it provides an equivalence relation between torsion and curvature. \cite{Harko-2014, Bahamonde1_2017}.

To study the universe more realistically, we consider it as an anisotropic fluid with viscosity, because  the bulk viscosity creates internal friction that converts the kinetic energy of particles into heat, an idea that could help to describe the recent acceleration of the universe. Therefore, bulk viscosity play a critical role in determining the dissipative effect of fluids. \cite{Brevik-2005, BrevikG-2005, Sadeghi_2013, Brevik-2017, Amaniali-2013}.

Since in physics it is believed that the nature of everything unknown to us is called dark or black, so the unknown parts of the universe are called dark energy and dark matter. Hence, we turn our attention to another dark part of the universe, i.e., dark matter. The nature of dark matter can be baryonic, this means, astronomical objects such as massive and compact haloes are composed of ordinary matter, but their electromagnetic radiation is negligible. Also, dark matter can be non-baryonic, that is, it is composed of hypothetical and real particles, in which axions and Weakly Interacting Massive Particles (WIMPs) are hypothetical particles, and Bose-Einstein condensate (BEC) is a state of matter created when particles called bosons are cooled to near absolute zero \cite{Mambrini-2016, Hooper-2007, Profumo-2003, Oikonomou-2007, Marsh-2016, Oikonomou-2022, OikonomouV-2022, Co-2020, CoR-2020, Barman-2021, Odintsov-2020, Oikonomou-2022, OikonomouV-2022, OdintsovSD-2019, OikonomouVK-2021, Harko1-2011, Das-2015, Fukuyama-2008, Li-2014, Das-2018, Boehmer-2007}. Hence dark matter can be described as a non-relativistic, Newtonian Bose-Einstein gravitational condensate gas, whose density and pressure are related by a barotropic equation of state \cite{Boehmer-2007}. For this purpose, we intend to relate dark matter and Bose-Einstein condensate (BEC). As we know, Bose-Einstein statistics are a concept that formed the basis of mathematics and enabled us to distinguish particular particles from each other. BEC is a substance in which dilute boson gas is cooled to very low temperatures. Due to this very low temperature, phase transfer occurs and most of the boson gases occupy the lowest quantum state, and the macroscopic quantum phenomenon appears at this point. Cold bosons fall on each other, and superparticles appear that behave like microwaves. This means that the dark matter is assumed to be a bosonic gas below the critical temperature that forms BEC \cite{Harko1-2011, Das-2015, Fukuyama-2008, Li-2014, Boehmer-2007, Das-2018}. By using the generalized Gross-Pitaevskii equation, the EoS of dark matter as a barotropic fluid is obtained, i.e., matter pressure, $p_m$, is only a function of matter energy density, $\rho_m$. Note that the corresponding EoS has been introduced as a normal dark matter \cite{Harko1-2011}. Also, in the same of Ref. \cite{Harko1-2011}, by considering the dark matter halo in a quantum ground state, the EoS of dark matter was obtained as $p_m \varpropto \rho_m^2$. After that, we extended the BEC model to the Extended Bose-Einstein Condensation (EBEC) model as cited in Ref. \cite{Mahichi-2021}. This means that dark matter is simultaneously considered in both cases of the normal dark matter and dark matter halo. In fact, EBEC regime helps us to understand more accurately the various epochs of the universe.

In general, dark matter and dark energy have the nature of attraction and repulsion, respectively. This means that the attractive force in dark matter pulls matter inside, but the repulsive force of dark energy pushes it outside. Therefore, in this study, we attribute the viscous $f(T, B)$ gravity model to dark energy and the EBEC model to dark matter. This assessment motivates us to investigate the nature of the largest cosmic scales by viscous $f(T,B)$ gravity, and to study the nature of individual galaxies by EBEC.

This paper is organized as follows:

In Sec. \ref{II}, we present a foundation of $f(T, B)$ gravity in the flat-FRW metric by bulk viscosity fluid.
In Sec. \ref{III}, we will explore EBEC as dark matter, and write the dark matter EoS as the virial expansion.
In Sec. \ref{IV}, we reconstruct the $f(T,B)$ gravity in terms of the redshift parameter, and fit the corresponding system by the Hubble data constraints.
In Sec. \ref{V}, we obtain the relevant cosmological parameters in terms of the redshift parameter and then plot the corresponding graphs.
In Sec. \ref{VI}, we analyse the stability and instability of the present model by the sound speed parameter.
Finally, in Sec. \ref{VII}, we provide a summary of the current job.

\section{Foundation of $f (T, B)$ theory}\label{II}

In this Sec., we intend to explore the modified teleparallel gravity entitled $f(T, B)$ model. The geometry of the teleparallel theory is raised as the dynamical variables entitled the components of tetrad field $e_{a} (x^{\mu } )$, which are introduced in the form of an orthonormal basis for the tangent space at each point $x^{\mu } $ of the space-time manifold. The corresponding metric in terms of tetrad field is written as $g_{\mu \nu } =\eta _{AB} e_{{\kern 1pt} {\kern 1pt} {\kern 1pt} \mu }^{A} e_{{\kern 1pt} {\kern 1pt} {\kern 1pt} \nu }^{B} $ that $\eta _{AB} =diag(-1,1,1,1)$ and $\det (e^{A} {}_{\mu } )=e=\sqrt{-g} $, in which the Greek letters and the Latin alphabets run over from 0 to 3, and denote space-time components and tangent space components, respectively. In that case, we write down $f(T, B)$ model as the following action
\begin{equation}\label{action1}
S=\int d^{4}  x ~e \left(\frac{f(T,B)}{\kappa ^{2} } +{\rm {\mathcal L}}_{m} \right),
\end{equation}
where $e$ and $\mathcal{L}_m$ are determinant of tetrad components and lagrangian of matter, respectively, as well as we have $\kappa^2 = 8 \pi G$ and $B=2 \partial_{\mu } (eT^{\mu } )/e$, in which $T^{\mu } $ as the torsion vector that can be defined by $T_{\mu } = T_{\nu \mu }^{\nu } $. We obtain Einstein field equation by making the variation of the action with respect to tetrad field as
\begin{eqnarray}\label{Ein1}
&{2e\delta _{\nu }^{\lambda } \nabla ^{\mu } \nabla _{\mu } \partial _{B} f-2e\nabla ^{\lambda } \nabla _{\nu } \partial _{B} f+eB\partial _{B} f\delta _{\nu }^{\lambda } +4e\left(\partial _{\mu } \partial _{B} f+\partial _{\mu } \partial _{T} f\right)S_{\nu } {}^{\mu \lambda } } \notag\\
&{+4e_{\nu }^{a} \partial _{\mu } \left(eS_{a} {}^{\mu \lambda } \right)\partial _{T} f-4e\partial _{T} fT^{\sigma } {}_{\mu \nu } S_{\sigma } {}^{\lambda \mu } -ef\delta _{\nu }^{\lambda } =16\pi e{\rm {\mathcal T}}_{\nu }^{\lambda } ,}
\end{eqnarray}
where ${\rm {\mathcal T}}_{\nu }^{\lambda } =e_{\nu }^{a} {\rm {\mathcal T}}_{a}^{\lambda } $ is the energy-momentum tensor of the matter. In the present work, we consider the flat-Friedmann-Robertson-Walker (FRW) metric as
\begin{equation}\label{Fried1}
ds^{2} =-dt^{2} +a^{2} (t)\left(dx^{2} +dy^{2} +dz^{2} \right),
\end{equation}
where $a(t)$ is the scale factor. The components of tetrad field are $e^{a} {}_{\mu } =diag(1,a,a,a)$. The torsion scalar, the torsion tensor, the asymmetry tensor, and the contorsion tensor are written, respectively, as
\begin{eqnarray}\label{tensors1}
&T=S_{\rho } {}^{\mu \nu } T^{\rho } {}_{\mu \nu }, \\
&T^{\rho } {}_{\mu \nu } = e_{A} {}^{\rho } \left(\partial _{\mu } e^{A} {}_{\nu } -\partial _{\nu } e^{A} {}_{\mu } \right),\\
&S_{\rho }^{\mu \nu } =\frac{1}{2} \left(K_{\rho }^{\mu \nu } +\delta _{\rho }^{\mu } T_{\alpha }^{\alpha \nu } -\delta _{\rho }^{\nu } T_{\alpha }^{\alpha \mu } \right),\\
&K_{\rho }^{\mu \nu } =-\frac{1}{2} \left(T^{\mu \nu } {}_{\rho } -T_{\rho }^{\nu \mu } -T_{\rho }^{\mu \nu } \right).
\end{eqnarray}

Now we can obtain the torsion scalar and the boundary term, respectively, in the following form
\begin{eqnarray}\label{TB1}
&T=6H^{2} ,\\
&B=6\left(\dot{H}+3H^{2} \right),
\end{eqnarray}
where $H=\dot{a}/a$ is Hubble parameter. It should be noted that standard action for general relativity is included only with a term of Ricci scalar, R, but in the modified teleparallel action is contained with terms of torsion scalar and boundary. These two theories (general relativity theory and modified teleparallel theory) have a difference only in a boundary term, i.e., $R=B - T=6\dot{H}+12H^{2} $. This issue shows us that the $f(T, B)$ gravity is an appropriate alternative for $f(R)$ gravity \cite{Bahamonde-2017, Bahamonde-2018}. Therefore, We obtain Friedmann equations by inserting Eq. \eqref{Fried1} into field equation \eqref{Ein1} as
\begin{subequations}\label{78}
\begin{eqnarray}
 &- 3{H^2}\left( {3{\partial _B}f + 2{\partial _T}f} \right) + 3H{\partial _B}\dot f - 3\dot H{\partial _B}f + \frac{1}{2}f = {\kappa ^2}{\mathcal{T}_0^0},\label{7}\\
 &- \left( {3{H^2} + \dot H} \right)\left( {3{\partial _B}f + 2{\partial _T}f} \right) - 2H{\partial _T}\dot f + {\partial _B}\ddot f + \frac{1}{2}\,\,f =  {\kappa ^2} \mathcal{T}_i^i,\label{8}
 \end{eqnarray}
\end{subequations}
where indices 0 and $i$ indicate components of time and space in time-space geometric. To have a more realistic model for evolution of the universe, we consider the viscose fluid instead of perfect fluid. Therefore, the energy–momentum tensor is written in the presence of the bulk viscosity as
\begin{equation}\label{Tij1}
\mathcal{T}_i^j=(\rho_{tot} + p_{tot} + p_{bulk})u_i u^j - \left(p_{tot} + p_{bulk}\right)\,  \delta_i^j,
\end{equation}
where $\rho_{tot}$, $p_{tot}$, and $p_{bulk} = -3 \xi H$ are the total energy density, the total pressure, and the pressure of bulk viscosity fluid inside the universe, respectively, in which $\xi$ is a positive constant for bulk viscosity. Also, the $4$-velocity is $u_\mu$ is $u^i$ = (+1,0,0,0) and have $ u_i u^j$ = 1. However, elements of energy--momentum tensor are as
\begin{subequations}\label{tau2}
\begin{eqnarray}
& \mathcal{T}_0^0 = \rho_{tot},\label{tau2-1} \\
& \mathcal{T}_i^i = -p_{tot} + 3 \xi H, i = 1,2,3.\label{tau2-2}
\end{eqnarray}
\end{subequations}

The total continuity equation in the presence of bulk viscosity yields
\begin{equation}\label{9}
{\dot \rho_{tot}} + 3 H ({\rho _{tot}} + {\overline{p}_{tot}}) = 0,
\end{equation}
where $\overline{p}_{tot}= {p_{tot}}-3 \xi H$. Now we consider that the contains of the inside universe contain both matter and torsion, then we have
\begin{subequations}\label{1011}
\begin{eqnarray}
 &{\rho _{tot}} = {\rho _m} + {\rho _{TB}},\label{10}\\
 &{\overline{p}_{tot}} = {p_m} + {p_{TB}}-3 \xi H,\label{11}
 \end{eqnarray}
\end{subequations}
where $\rho_{TB}$ and $p_{TB}$ are the energy density and the pressure for dark energy, respectively. Therefore, we can immediately write continuity equations of matter and torsion in the following form
\begin{subequations}\label{conteq1}
\begin{eqnarray}
 &{\dot \rho_{m}} + 3 H ({\rho_{m}} + {{p}_{m}}) = Q,\label{conteq1-1}\\
 &{\dot \rho_{TB}} + 3 H ({\rho _{TB}} + {{p}_{TB}} - 3 \xi H) = -Q,\label{conteq1-2}
 \end{eqnarray}
\end{subequations}
where $Q$ is an interaction term between the elements of the universe. When energy flow is transferred from dark matter to dark energy, the interaction term becomes a negative, and when energy flow is transferred from dark energy to dark matter, the interaction term becomes a positive. From the perspective of dimensionality, the $Q$ term is equal to the product of the Hubble parameter and the energy density. Therefore, we take $Q = 3 b^2 H \rho_m$ in which $0< b < 1$ is the intensity of energy transfer \cite{Setare-2009, Chimento-2003, Guo-2005}.

In that case, we write down $\rho _{TB}$ and $p _{TB}$ by using of Eqs. \eqref{78}, \eqref{tau2}, and \eqref{1011} as below
\begin{subequations}\label{1213}
\begin{eqnarray}
 &{\rho _{TB}} = \frac{1}{{{\kappa ^2}}}\left( { - 3{H^2}\left( {3{\partial _B}f + 2{\partial _T}f} \right) + 3H{\partial _B}\dot f - 3\dot H{\partial _B}f + \frac{1}{2}f} \right) - {\rho _m},\label{12}\\
 &{p}_{TB} = \frac{{ - 1}}{{{\kappa ^2}}}\left( { - \left( {3{H^2} + \dot H} \right)\left( {3{\partial _B}f + 2{\partial _T}f} \right) - 2H{\partial _T}\dot f + {\partial _B}\ddot f + \frac{1}{2}f} \right) - {p_m} + 3 \xi H.\label{13}
 \end{eqnarray}
\end{subequations}

The equation of state (EoS) for the dark energy is
\begin{equation}\label{14}
\omega _{TB} = \frac{p_{TB}}{\rho _{TB}},
\end{equation}
where the EoS is dependent on the parameters of $f(T,B)$ model and the viscous fluid.

\section{The extended Bose-Einstein condensate as dark matter}\label{III}

The Bose-Einstein condensation, known as the "fifth state of matter," is a state of matter that occurs when boson particles cool to near zero (minus 273.15 degrees Celsius). At such a low temperature, the particles do not have enough energy to enter situations where their distinct quantum properties overlap. Without the energy difference to separate the particles, they all merge into a single quantum state to form superparticles that behave like a microwave, unlike ordinary particles. The speed of light passing through the Bose-Einstein condensate is very slow. According to Bose and Einstein's experiments and observations, when atoms approach absolute zero temperature, the waves expand and eventually overlap, so that their constituent elementary particles all merge into a single quantum state, which this condition is called Bose-Einstein condensation. Here we are going to describe one of the components of the universe called dark matter with the help of Extended Bose-Einstein Condensation (EBEC). In fact, the described model can be similar to the model of our universe. In order to obtain the EoS of dark matter, we take BEC description from the generalized Gross-Pitaevskii equation that leads to developing it in cosmology \cite{Pitaevskii-2003, Pethick-2008}. Thus, EoS of dark matter is written as
\begin{equation}
p_{m} =u_{0}~ \rho_{m}^{2},
\end{equation}
where $u_{0} =2\pi l_{a} /m_{m}^{3} $, in which $l_{a} $ is the s-wave scattering length (refer to Ref. \cite{Harko-2011, Chavanis-2017, Harko-2015} for more details). As mentioned above, we study dark matter from the perspective of Bose-Einstein condensation, and also use EBEC regime to better understand the various periods of the universe from the early time to the late time. Then write the equation of state for dark matter as follows:
\begin{equation}\label{pressure3}
p_m = \alpha \rho_{m} + \beta \rho_{m}^2,
\end{equation}
where $\alpha$ is introduced as single-body interaction, resulting from normal dark matter and $\beta$ is introduced as two-body interaction, resulting from dark matter halo. Because $\alpha$ and $\beta$ plays an important role in understanding the nature of dark matter in the universe, the values of $\alpha$ and $\beta$ indicate the both contribution of the normal dark matter and dark matter halo \cite{Mahichi-2021}.

By looking closely at Eqs. \eqref{pressure3}, we find that this equation is the EoS of the viral expansion, which in fact helps us to write the pressure of dark matter as a power series in terms of the energy density of dark matter.
\begin{equation}\label{pressure4}
p_{m} = \sum a_i \rho^i_{m},
\end{equation}
where the first term, $a_1 = \alpha$, represents normal dark matter, the second term, $a_2 = \beta$, represents the dark matter halo in a quantum ground state, and the third term, $a_3$, and higher terms, $a_i$, are related to the excited quantum system of the dark matter halo. Since we deal with forms of barotropic and halo for dark matter in this work, so we only take the first two terms of the virial expansion. In that case, we look forward to the above consequences, which will help us to understand the early era to the late era of the universe.

In order to obtain the energy density of dark matter, we insert Eq. \eqref{pressure3} into \eqref{conteq1-1} and have
\begin{equation}\label{rhoch1}
\rho_{m} = \rho_0\, \frac{c\, \eta \, a_0^{3 \eta} - \beta}{c\, \eta \, a^{3 \eta} - \beta},
\end{equation}
where $\rho_0 = \eta /(c\, \eta\, a_0^{3 \eta} - \beta)$, $a_0$, and $c$ are respectively the present energy density of dark matter, the current scale factor, and an integral constant, in which $\eta = \alpha + 1 - b^2$. It should be noted that this relationship shows that it depends on the scale factor, interacting term, and coefficients of EBEC dark matter.


\section{Reconstruction of $f(T,B)$ gravity, and the Hubble data constraints}\label{IV}

In this section, we intend to reconstruct the corresponding Friedman equation according to the redshift parameter so that we can solve the corresponding Friedman equations in the presence of EBEC for $f(T, B)$ gravity. For this purpose, we write the relationship between redshift parameter, $z$, and the present scale factor, $a_{0}$, as $1+z = a_{0}/a$. Also, we can clearly write the relationship between the time derivative and the redshift derivative in the form of $d/dt=-H(1+z)\, d/dz$. In order to solve the present system for interacting $f(T, B)$ gravity in presence of EBEC, we introduce the below relation
\begin{equation}\label{H2H0}
H^2 = H_0^2 E(z),
\end{equation}
where $E(z)$ is the parametrization function that is functional of the redshift parameter, and the late time Hubble parameter is equal to $H_{0} = 67.4 \pm 0.5 \, km \, s^{-1} \, Mpc^{-1}$ \cite{Aghanim-2020}. In that case, torsion scalar, $T$, and boundary term, $B$, are written in terms of redshift parameter as
\begin{eqnarray}
&T=6 H_{0}^{2} E(z),\\
&B=3 H_{0}^{2} (1+z) E'(z) - 18 H_{0}^{2}  E(z),
\end{eqnarray}
where the prime index displays the derivative with respect to redshift parameter. The energy density and the pressure of dark energy in Eqs. \eqref{1213} are rewritten in terms of redshift parameter as follows:
\begin{subequations}\label{rhopTB1}
\begin{eqnarray}
&\rho _{TB} =\frac{1}{\kappa ^{2}} \Big[-3H_{0}^{2} E\left(3\partial _{B} f+2\partial _{T} f\right)-3H_{0}^{2} (1+z)E\partial _{B} f'+\frac{3}{2} H_{0}^{2} (1+z)E'\, \partial _{B} f \notag \\
&+\frac{1}{2} f\Big] - \rho _{m},\\
&p_{TB} =\frac{1}{\kappa ^{2} } \Big[H_{0}^{2} \left(3E-\frac{1}{2} \left(1+z\right)E'\right)\left(3\partial _{B} f+2\partial _{T} f\right)-2H_{0}^{2} (1+z)E\, \partial _{T} f' -\frac{1}{2} H_{0}^{2} (1+z)^{2} E'\, \partial_{B} f' \notag \\
&-\frac{1}{2} f - H_{0}^{2} (1+z)E\, \partial_{B} f'-H_{0}^{2} (1+z)^{2} E\, \partial _{B} f''\Big] - \alpha \rho_{m} - \beta \rho_m^2 + 3 \xi H_{0} \sqrt{E}.
\end{eqnarray}
\end{subequations}

In order to solve the corresponding system, we use a model that is commonly used for simplicity and ease of calculations. For this purpose, we consider a power-law model for the scale factor in terms of cosmic time as follows:
\begin{equation}\label{at1}
a(t) = a_{0} \left(\frac{t}{t_{0}} \right)^{s},
\end{equation}
where $a_{0}$ is the current scale factor, and $s$ is a dimensionless positive coefficient as a correction factor \cite{Tutusaus-2016}. The Hubble parameter easily yields
\begin{equation}\label{hubpar1}
H = \frac{s}{t},
\end{equation}
where in the late time we will have
\begin{equation}\label{hubpar2}
t_0 = \frac{s}{H_0},
\end{equation}
where $t_0$ is the age of the universe\footnote{Note that the age of the universe is $t_0 = \frac{1}{H_0} \int_0^{\infty} \,\frac{dz}{(1+z) \sqrt{E}}$ which is the same as the result of Eq. \eqref{hubpar2}}. Also, we earn the Hubble parameter in terms of redshift parameter as
\begin{equation}\label{H2H01}
H^{2}(z) = H_{0}^{2} \left(1+z\right)^{\frac{2}{s}},
\end{equation}
where
\begin{equation}\label{H2H02}
E(z)=\left(1+z\right)^{\frac{2}{s}}.
\end{equation}

Note that Eq. \eqref{H2H01} has only one free parameter, $s$, which can easily be examined with observational data. To demonstrate the accuracy of our model, we fit it with 53 supernova data for Hubble parameter that shown in Tab. \ref{tab1}, data sets collected from Refs. \cite{Zhang14, Jimenez03, Simon05, Moresco12, Gaztanaga09, Oka14, Wang17, Chuang13, Alam17, Moresco16, Ratsimbazafy17, Anderson-2014, Blake12, Stern10, Moresco15, Busca13, Bautista17, Delubac15, FontRibera14}. The estimate is in the range of $0.07 \leq z \leq 2.36$, in which the corresponding data sets are measured by techniques of galaxy differential age or cosmic chronometer and radial BAO size methods. Since the present work related to statistics estimation, then we take maximum likelihood analysis with chi-squared value, $\chi^2_{min}$, which is introduced as a very essential instrument for fitting of the cosmological data. The chi-squared is
\begin{equation} \label{likelihood1}
\chi_{min}^2 = \sum_{i = 1}^{53} \frac{\left(H_{obs}(z_i) - H_{th} (z_i, H_0)\right)^2}{\sigma_H^2 (z_i)},
\end{equation}
where $H_{obs}$ and $H_{th}$ display the observed value and the theoretical value of the Hubble parameter data, respectively, and $\sigma_H$ represents the standard error or uncertainty values in the observed values. After examining with the Hubble parameter data set and minimizing $\chi^2_{min}$, we obtain $s = 0.95$ by the best fitting. Also, we can earn the age of the universe since the Big Bang as $t_0 = 13.78 Gyr$ from Eq. \eqref{hubpar2}. Therefore, the obtained age of the universe is compatible with $\Lambda$-CDM model \cite{Aghanim-2020}, which indicates the validity of the present model.

\begin{table}[h]
\caption{The Hubble parameter data set in terms of redshift parameter and their uncertainty values, units of $H(z)$ and $\sigma_H$ are $km~ s^{-1} ~Mpc^{-1}$} 
\centering 
\begin{tabular}{||c | c | c | c | c || c | c | c | c | c||} 
\hline\hline 
No. & Redshift & H(z) & $\sigma_{H}$ & Ref. & No. & Redshift & H(z) & $\sigma_{H}$ & Ref.\\ [0.5ex] 
\hline 
1. & 0.07 & 69.0 & 19.6 & \cite{Zhang14} & 28. & 0.510 & 90.4 & 1.9 &  \cite{Alam17}\\
2. & 0.09 & 69.0 & 12.0 & \cite{Jimenez03} & 29. & 0.52 & 94.35 & 2.64 &  \cite{Wang17}\\
3. & 0.12 & 68.6 & 26.2 & \cite{Zhang14} & 30. & 0.56 & 93.34 & 2.3 & \cite{Wang17}\\
4. & 0.17 & 83 & 8 &  \cite{Simon05}  & 31. & 0.57 & 92.9 & 7.855 &  \cite{Anderson-2014}\\
5. & 0.179 & 75.0 & 4.0 & \cite{Moresco12} & 32. & 0.59 & 98.48 & 3.18 &  \cite{Wang17}\\
6. & 0.199 & 75.0 & 5.0 &  \cite{Moresco12} & 33. & 0.593 & 104.0 & 13.0 & \cite{Moresco12}\\
7. & 0.200 & 72.9 & 29.6 &  \cite{Zhang14} & 34. & 0.6 & 87.9 & 6.1 &  \cite{Blake12} \\
8. & 0.24 & 79.69 & 3.32 &  \cite{Gaztanaga09} & 35. & 0.610 & 97.3 & 2.1 &  \cite{Alam17} \\
9. & 0.27 & 77 & 14 &  \cite{Simon05}  & 36. & 0.64 & 98.02 & 2.98 &  \cite{Wang17}\\
10. & 0.280 & 88.8 & 36.6 & \cite{Zhang14} & 37. & 0.680 & 92.0 & 8.0 & \cite{Moresco12}\\
11. & 0.30 & 81.7 & 5.0 &  \cite{Oka14}  & 38. & 0.73 & 97.3 & 7 &  \cite{Blake12}\\
12. & 0.31 & 78.18 & 4.74 &  \cite{Wang17} & 39. & 0.781 & 105.0 & 12 &  \cite{Moresco12}\\
13. & 0.34 & 83.8 & 2.96 &  \cite{Gaztanaga09} & 40. & 0.875 & 125 & 17 &  \cite{Moresco12}\\
14. & 0.35 & 82.7 & 9.1 &  \cite{Chuang13} & 41. & 0.880 & 90.0 & 40.0 &  \cite{Stern10}\\
15. & 0.352 & 83 & 14 &  \cite{Moresco12} & 42. & 0.900 & 117 & 23.0 &  \cite{Simon05}\\
16. & 0.36 & 79.94 & 3.38 &  \cite{Wang17} & 43. & 1.037 & 154 & 20 &  \cite{Moresco12}\\
17. & 0.38 & 81.5 & 1.9 &  \cite{Alam17} & 44. & 1.300 & 168 & 17 &  \cite{Simon05}\\
18. & 0.3802 & 83.0 & 13.5 &  \cite{Moresco16} & 45. & 1.363 & 160 & 33.6 & \cite{Moresco15}\\
19. & 0.40 & 82.04 & 2.03 &  \cite{Wang17} & 46. & 1.430 & 177 & 18 &  \cite{Simon05}\\
20. & 0.4004 & 77 & 10.2 &  \cite{Moresco16} & 47. & 1.530 & 140 & 14 &  \cite{Simon05}\\
21. & 0.4247 & 87.1 & 11.2 &  \cite{Moresco16} & 48. & 1.750 & 202 & 40 &  \cite{Simon05}\\
22. & 0.43 & 86.45 & 3.27 &  \cite{Gaztanaga09} & 49. & 1.965 & 186.5 & 50.4 & \cite{Moresco15}\\
23. & 0.44 & 84.81 & 1.83 &  \cite{Wang17} & 50. & 2.30 & 224 & 8.6 &  \cite{Busca13}\\
24. & 0.4497 & 92.8 & 12.9 &  \cite{Moresco16} & 51. & 2.33 & 224 & 8 & \cite{Bautista17}\\
25. & 0.470 & 89 & 34 & \cite{Ratsimbazafy17} & 52. & 2.340 & 222 & 7 &  \cite{Delubac15}\\
26. & 0.4783 & 80.9 & 9.0 &  \cite{Moresco16} & 53. & 2.360 & 226 & 8 &  \cite{FontRibera14}\\
27. & 0.48 & 87.79 & 2.03 &  \cite{Wang17} &  &  &  &  &  \\
[1ex] 
\hline 
\end{tabular}
\label{tab1} 
\end{table}

In the next section, we will investigate the obtained results from the Hubble parameter data set constraints in the presence of EBEC for $f(T, B)$ gravity.

\section{The interaction between viscous $f(T,B)$ gravity and EBEC}\label{V}

In this section, we intend to explore the interaction between EBEC and $f(T, B)$ gravity in the presence of bulk viscosity. Therefore, in order to compute the present model, the form of function $f(T, B)$ plays a very essential role. To do this, we have to opt a viable cosmic model for $f(T, B)$ gravity that can provide information about the accelerating universe in the late era. One of these options is mixed power-law model, which is written for the function $f(T, B)$ as follows:
\begin{equation}\label{fTB2}
f(T,B) = \lambda B^{m} T^{n} ,
\end{equation}
where $\lambda$, $m$, and $n$ are constant coefficients. According to this proposal, the contribution of torsion term or boundary term may occur separately or together.

Now by replacing Eqs. \eqref{rhoch1}, \eqref{H2H02}, and \eqref{fTB2} into Eqs. \eqref{rhopTB1} we obtain the energy density and the pressure of dark energy versus the redshift parameter in the following form
\begin{subequations}\label{rhopTB2}
\begin{eqnarray}
&\rho _{TB} =\frac{ 6^{m+n} \lambda \delta H_0^{2(m+n)} (3 s - 1)^{m-1}}{2 \kappa ^{2} s^m} (1+z)^{\frac{2}{s} (m+n)} - \frac{\eta}{c \eta \left(\frac{a_0}{1+z}\right)^{3 \eta} - \beta},\label{rhopTB2-1}\\
&p_{TB} =\frac{ 6^{m+n-1} \lambda \delta (2 m+2 n -3s) H_0^{2(m+n)} (3 s - 1)^{m-1}}{\kappa ^{2} s^{m+1}} (1+z)^{\frac{2}{s} (m+n)} - \frac{\alpha \eta}{c \eta \left(\frac{a_0}{1+z}\right)^{3 \eta} - \beta} \notag \\
&- \frac{\beta \eta^2}{\left(c \eta \left(\frac{a_0}{1+z}\right)^{3 \eta} - \beta\right)^2} +3 \xi H_{0} (1+z)^{\frac{1}{s}},\label{rhopTB2-2}
\end{eqnarray}
\end{subequations}
where $\delta = 3 m + 2 n + 3 s - 2 m^2 - 2 m n - 3 m s - 6 n s - 1$.

In order to get an acceptable answer from the above calculations, at least one set of free parameter values should be selected so that the results represent an expanding universe. This means that the free parameters play a very important role in this model. Therefore, our choice is motivated by the fact that the energy density and pressure of dark energy become positive and negative, respectively, and also lead to the crossing of the divided-phantom line in the variations of EoS versus the redshift parameter. Therefore, a series of selected values are: $\lambda = 75$, $c = 0.001$, $b = 0.1$, $m = 1.75$, $n = -1.5$, $\alpha = 3$, $\beta = 0.0012$, $a_0 = 0.75$, and $\xi = 0.1, 5$. Note that the free parameters $\lambda$, $c$, $b$, $m$, $n$, and $a_0$ are dimensionless, and $\xi$ has units $pa.s$ in SI system or $M^3$ in Planck systems.

\begin{figure}[ht]
\begin{center}
{\includegraphics[scale=.35]{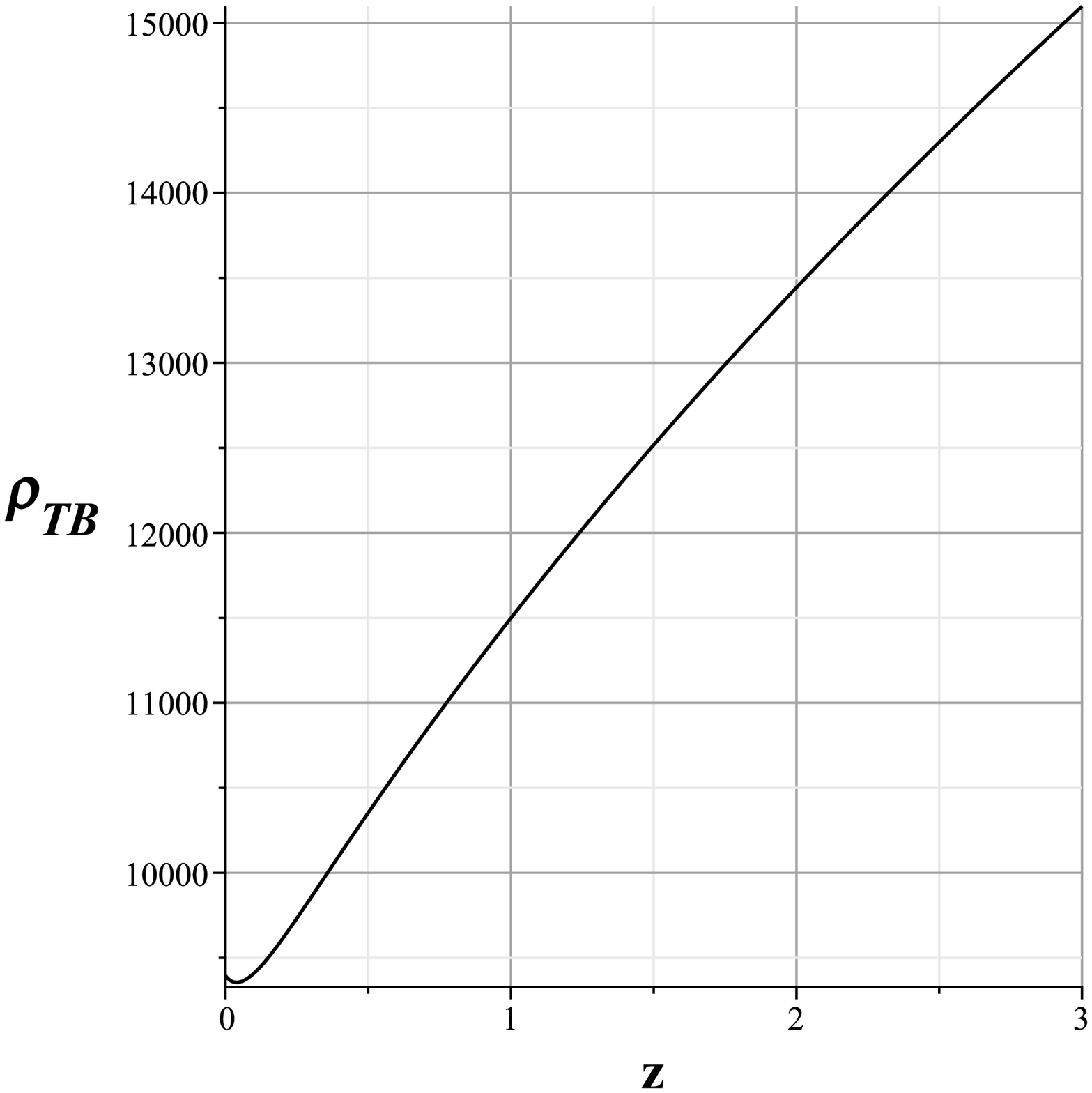}}
{\includegraphics[scale=.35]{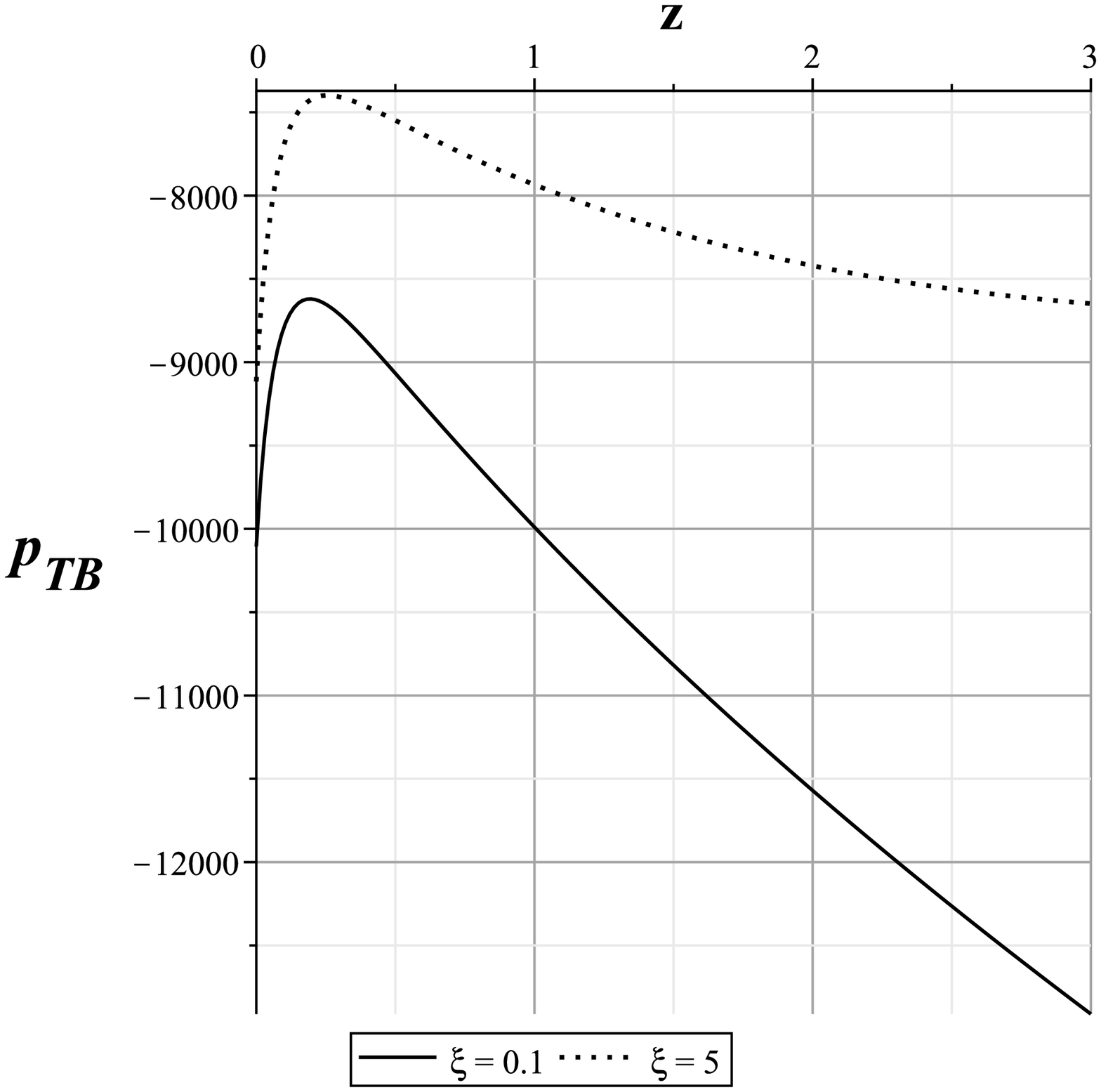}}
\caption{The energy density and the pressure of dark energy in terms of redshift parameter.}\label{figurerhop}
\end{center}
\end{figure}

Now we plot the variation of $\rho_{TB}$ and $p_{TB}$ versus to redshift parameter as shown in Figs. \ref{figurerhop}. Therefore, the variation of $\rho_{TB}$ depends on the coefficients of $F(T, B)$ gravity, the interacting model, and EBEC dark matter, but, the variation of $p_{TB}$ is also dependent on bulk viscosity in addition to the mentioned ones. Also, it is evident that the value of $\rho_{TB}$ is more than zero, and the value of $p_{TB}$ is less than zero for the present universe ($z=0$). It should be noted that $\rho_{TB}$ and $p_{TB}$ have the same unit $M L^{-1} T^{-2}$ in $SI$ system and or units $M^4$ in Planck system.

\begin{figure}[ht]
\begin{center}
{\includegraphics[scale=.35]{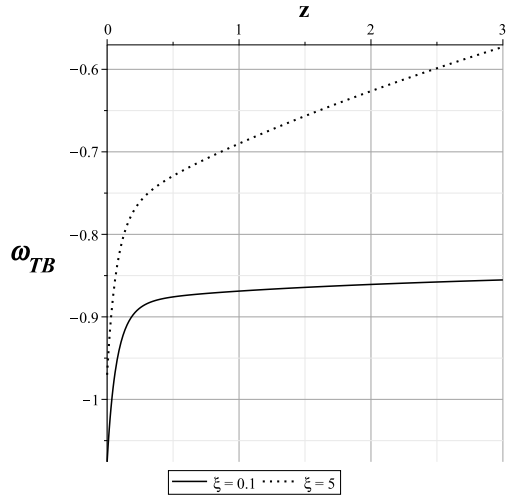}}
\caption{EoS parameter of dark energy in terms of redshift parameter.}\label{figureomega}
\end{center}
\end{figure}

On the other hand, by inserting Eqs. \eqref{rhopTB2} into Eq. \eqref{14}, we can calculate EoS of dark energy, $\omega_{TB}$, as
\begin{equation}\label{omegaTB2}
\omega_{TB} = \frac{\frac{(2 m+2 n-3 s)}{3 s} A (1+z)^{\frac{2}{s}(m+n)} - \alpha \rho_m - \beta \rho_m^2 + 3 \xi H_0 (1+z)^{\frac{1}{s}}}{A  (1+z)^{\frac{2}{s}(m+n)} - \alpha \rho_m},
\end{equation}
where $A = 6^{m+n} \lambda \delta H_0^{2(m+n)} (3 s - 1)^{m-1} / (2 \kappa ^{2} s^m)$. The variation of EoS of dark energy is plotted versus redshift parameter as shown in Fig. \ref{figureomega}.

As we know, EoS parameter plays an important role in modern cosmology which is a dimensionless quantity. So that,  this parameter is a very good description for different eras of the formation of the universe, i.e, we will have eras of accelerated phase (quintessence phase), cosmological constant, and phantom phase for $-1 < \omega < -\frac{1}{3}$, $\omega = -1$ and $\omega < -1$, respectively. However the corresponding graph \ref{figureomega} represents us that the formation of the universe enters the era of the accelerated phase and crosses the divided-phantom line. Therefore, the present values of EoS for $\xi = 0.1$ and $\xi = 5$ are $-1.075$ and $-0.97$, respectively. In that case, we can observe that the universe moves between regions quintessence and phantom when the bulk viscosity coefficient varies. It is worth noting that the values of the dark energy EoS parameter versus the redshift parameter vary as shown in Fig. \ref{figureomega}. Ref. \cite{Aghanim-2020} is a report for Planck 2018 results measuring the dark energy EoS parameter as $-1.03 \pm 0.03$ with Type Ia supernovae (SNe), and earlier in 2010, Ref. \cite{Amanullah_2010} provided the best fit for the dark energy EoS parameter to Union2 data in flat-universe as $-0.997^{+0.050}_{-0.054}$ for statistical uncertainty and $-0.997^{+0.077}_{-0.082}$ for statistical  and systematic uncertainties. Also, Ref. \cite{Scolnic_2018} measured the dark energy EoS parameter as $-1.026 \pm 0.041$ with the Pantheon Sample. According to the above evidence, the universe is undergoing accelerated expansion, which is consistent with the results obtained in Ref. \cite{Aghanim-2020, Amanullah_2010, Scolnic_2018} and even consistent with a cosmological constant.

\section{Stability and instability analysis}\label{VI}

\begin{figure}[t]
\begin{center}
{\includegraphics[scale=.35]{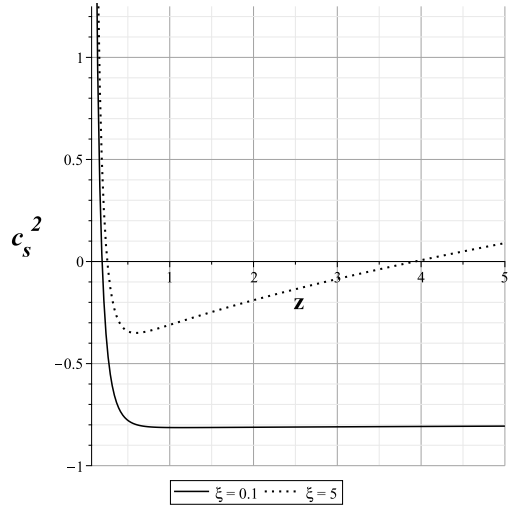}}
\caption{The sound speed parameter of dark energy in terms of redshift parameter.}\label{cs1}
\end{center}
\end{figure}

In this section, we intend to explore the stability and the instability of the our model from a thermodynamic point of view based on the sound speed parameter, $c_s^2$. From a thermodynamic perspective, the universe is in an adiabatic system, in such a way that no energy or mass is exchanged from the universe to the outside. In that case, the entropy perturbation is zero, then we have
\begin{equation}\label{cs21}
\delta p_{TB} (S, \rho_{TB}) = \frac{\partial p_{TB}}{\partial S} \delta S + \frac{\partial p_{TB}}{\partial \rho_{TB}} \delta \rho_{TB} = c_s^2 \delta \rho_{TB},
\end{equation}
where $c_s^2 = \partial_z p_{TB} / \partial_z \rho_{TB}$, and symbol $\partial_z$ represents derivative with respect to redshift parameter. Now the values of the sound speed parameter play a key role in describing the stability and instability of the system. Therefore, the corresponding parameter has the conditions $c_s^2 > 0$ and $c_s^2 < 0$ which respectively declare the stability and instability of the model. We draw the variation of the sound speed parameter in terms of redshift parameter as shown in Fig. \ref{cs1}.
We can see from Fig. \ref{cs1} for $\xi = 0.1$ that the formation of the universe begins instability phase and continues to the stability phase at the late time. Also in $\xi = 5$, we see that the formation of the universe begins from a stability phase and then at the threshold of the accelerated phase enters an instability phase, and again return to stability phase in late time. Therefore, with the existence of viscous fluid gives rise to the different stability analyses for the formation of the universe. At the end, it can be said that the phases of stability and instability indicate that the energy density is or is not in a controllable growth, respectively.
\begin{figure}[h]
\begin{center}
{\includegraphics[scale=.35]{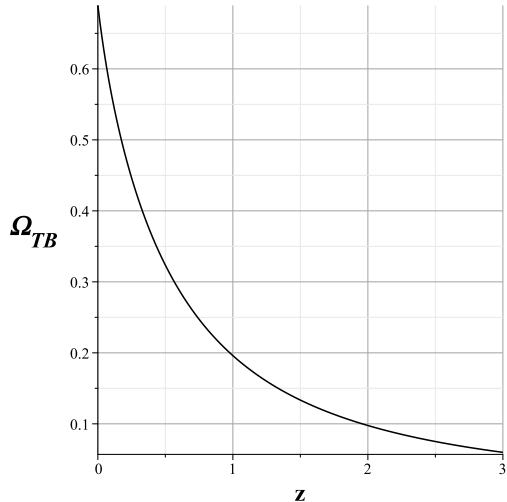}}
\caption{The density parameter of dark energy in terms of redshift parameter.}\label{Omega2}
\end{center}
\end{figure}

Now, on the other hand, we are looking for more complete explanations for this model. For this purpose, we estimate the density parameter of dark energy, $\Omega_{TB} = \rho_{TB} / \rho_c$, where $\rho_c = 3 H^2 / \kappa^2$ is the critical density. By substituting \eqref{rhopTB2-1} into $\Omega_{TB}$, we can plot the variation of the density parameter in terms of redshift parameter as shown in Fig. \ref{Omega2}. We can see from Fig. \ref{Omega2}, that the amount of $\Omega_{TB}$ increases from the early universe to the late universe, in which the present amount of the density parameter of dark energy is $0.69$. According to the Planck 2018 results report \cite{Aghanim-2020}, the value of the dark energy density parameter with Type Ia supernovae (SNe) has been obtained as $0.6847 \pm 0.0073$, so that it is compatible with our results. As a result, the contribution of dark energy is increasing, which indeed indicates that the fate of the universe will continue to expand.

\section{Conclusion}\label{VII}

In this paper, we studied the dark parts of the universe by $f(T, B)$ gravity and EBEC. First we explored the modified teleparallel gravity entitled $f(T, B)$ model in a viscous fluid by the flat-FRW universe. Then, by writing the Einstein equation, we earned the Friedmann equations in the presence of bulk viscosity. Also, we considered an interaction between the contents of the universe, i.e., dark matter and dark energy with term of $Q= 3 b^2 H \rho_m$ for a more realistic universe. Next, the continuity equations, and EoS parameter of dark energy have been written in terms of matter and torsion. The interesting point of $f(T,B)$ gravity model is that boundary term, $B$ relates to the Ricci scalar $B = R + T$, i.e., $f(T,B)$ gravity can covers the gravity models of $f(T)$ based on the Weitzenb\"{o}ck connection and the $f(R)$ based on the Levi-Civita connection. In fact, the $f(T,B)$ model covers both models simultaneously.

On the other hand, we studied EBEC as an appropriate alternative instead of the dark matter. In fact, we describe BEC from the generalized Gross-Pitaeveskii equation and use EBEC regime to write the EoS of dark matter. The interesting point of EBEC regime is that the EoS of dark matter is adapted to the EoS of the virial expansion, which is written in terms of normal dark matter and dark matter halo. Therefore, the elegance of the EBEC regime gives a better understanding of the different epochs of the universe from the early time to late time.

In what follows, in order to examine the present model with the astronomical data, we took the power-law for the scale factor and then wrote the Hubble parameter in terms of the redshift parameter. Next, the corresponding Hubble parameter was fitted with $53$ supernova data, so-called the Hubble data constraints, and as a result, the age of the universe  as $t_0 = 13.75 \,Gyr$ was found. After that, we opted a viable cosmic model for $f(T,B)$ gravity as mixed power-law model that presented information on the late universe. In addition, we reconstructed the energy density, the pressure, and the EoS of dark energy in terms of redshift parameter and some free parameters that come from $f(T, B)$ gravity, viscous fluid, interacting model, and EBEC regime. Note that choice of free parameters is very sensitive to prove that the universe has an expanding acceleration. In that case, we plotted $\rho_{TB}$, $p_{TB}$, and $\omega_{TB}$ in terms of redshift parameter for some acceptable free parameters. So that, the present amounts of the EoS parameter are equal to $-1.075$ and $-0.97$ for $\xi = 0.1$ and $\xi = 5$, respectively. Therefore, we concluded that the obtained results are completely consistent with the observational data.

For a more thorough review, we analyzed the stability and instability of the present model by the sound speed parameter. In that case, we drew the variation of the sound speed parameter in terms of the redshift parameter with different values of bulk viscosity coefficient. The results showed us that the stability conditions passed through a series of instability conditions and reached a stability condition at the late time. This means that the energy density of dark energy is in a controlled growth. Finally, in order to make the results of the present study consistent with the observational data, we calculated the value of the dark energy density parameter with a value of $0.69$, as shown in Fig. \ref{Omega2}. As a further work, we propose the development of the present work for axion models.

\section{Acknowledgment}

We thank Prof. Iver H. Brevik for his useful comments and suggestions to improve our work.


\end{document}